\documentclass[12pt,preprint]{aastex}
\usepackage{apjfonts}
\shorttitle{11 Gyr OLD WDs}
\shortauthors{KILIC ET AL.}
\begin{document}
\title{Visitors from the Halo: 11 Gyr old White Dwarfs in the Solar Neighborhood\footnote{Based
on observations obtained at the MMT Observatory, a joint facility of the Smithsonian Institution
and the University of Arizona.}}

\author{Mukremin Kilic\altaffilmark{1,8},
Jeffrey A. Munn\altaffilmark{2},
Kurtis A. Williams\altaffilmark{3},
P. M. Kowalski\altaffilmark{4,5},
Ted von Hippel\altaffilmark{6},
Hugh C. Harris\altaffilmark{2},
Elizabeth J. Jeffery\altaffilmark{7},
Steven DeGennaro\altaffilmark{3},
Warren R. Brown\altaffilmark{1}, and
B. McLeod\altaffilmark{1}}

\altaffiltext{1}{Smithsonian Astrophysical Observatory, 60 Garden Street, Cambridge, MA 02138; mkilic@cfa.harvard.edu}
\altaffiltext{2}{US Naval Observatory, P.O. Box 1149, Flagstaff, AZ 86002}
\altaffiltext{3}{Department of Astronomy, 1 University Station C1400, Austin, TX 78712}
\altaffiltext{4}{Helmholtz-Centre Potsdam - GFZ German Research Centre for Geosciences, D-14473, Potsdam, Germany}
\altaffiltext{5}{Lehrstuhl f\"ur Theoretische Chemie, Ruhr-Universit\"at Bochum, 44780 Bochum, Germany}
\altaffiltext{6}{Physics Department, Siena College, 515 Loudon Road, Loudonville, New York 12211}
\altaffiltext{7}{Space Telescope Science Institute, 3700 San Martin Drive, Baltimore, MD 21218}
\altaffiltext{8}{\em Spitzer Fellow}

\begin{abstract}

We report the discovery of three nearby old halo white dwarf candidates in the Sloan Digital Sky Survey (SDSS),
including two stars in a common proper motion binary system. These candidates are selected from our 2800 square
degree proper motion survey on the Bok and U.S. Naval Observatory Flagstaff Station 1.3m telescopes, and
they display proper motions of $0.4-0.5\arcsec$ yr$^{-1}$.
Follow-up MMT spectroscopy and near-infrared photometry demonstrate that all three objects are hydrogen-dominated atmosphere
white dwarfs with $T_{\rm eff} \approx 3700 - 4100$ K. For average mass white dwarfs, these temperature estimates
correspond to cooling ages of $9-10$ Gyr, distances of $70-80$ pc, and tangential velocities of $140-200$ km s$^{-1}$.
Based on the $UVW$ space velocities, we conclude that they most likely belong to the halo. Furthermore, the combined
main-sequence and white dwarf cooling ages are 10-11 Gyr. Along with SDSS J1102+4113, they
are the oldest field white dwarfs currently known. These three stars represent only a small fraction of
the halo white dwarf candidates in our proper motion survey, and they demonstrate that deep imaging surveys like
the Pan-STARRS and Large Synoptic Survey Telescope should find many old thick disk and halo white dwarfs that can be used
to constrain the age of the Galactic thick disk and halo.

\end{abstract}

\keywords{stars: atmospheres---stars: evolution---white dwarfs}

\section{Introduction}

White dwarf (WD) cosmochronology provides an independent and accurate age dating method
for different Galactic populations \citep{winget87,liebert88}. Using 43 cool WDs in
the Solar Neighborhood, \citet{leggett98} derived a disk age of 8 $\pm$ 1.5 Gyr. 
\citet{kilic06} and \citet{harris06} significantly improved the field WD sample
by using SDSS and USNO-B astrometry to select high proper motion candidates. However,
their survey suffered from the magnitude limit of the Palomar Observatory Sky Survey
plates and they were unable to find many thick disk or halo WD candidates.

Substantial investment of the $Hubble~Space~Telescope$ time on two globular clusters, M4 and NGC 6397,
revealed clean WD cooling sequences. \citet{hansen04,hansen07} and \citet{bedin09} use these data
to derive cooling ages of $\approx12$ Gyr for the two clusters. 
The coolest WDs in these clusters are about 650 $\pm$ 230 K cooler than
the coolest WDs in the disk \citep{kowalski07a}.
These studies demonstrate that the Galactic
halo is older than the disk by $\geq2$ Gyr \citep{hansen02,fontaine01,kowalski07a}. 
Even though the WDs in globular clusters provide reliable age estimates,
these clusters may not represent the full age range of the Galactic halo. The required exposure times to reach the WD
terminus in globular clusters limit these studies to the nearest few clusters. In addition, only two-filter
($V$ and $I$) photometry is used to model the absolute magnitude and color distribution of the oldest WDs to derive ages.
The far closer and brighter WDs of the local halo field are an enticing alternative as well as
complementary targets, with the additional potential to constrain the age
range of the Galactic halo. Accurate ages for field WDs can be obtained through optical
and near-infrared photometry and trigonometric parallax measurements.
Nearby WDs can also be used to understand the model uncertainties and put the Globular
cluster ages on a more secure footing.

The quest for field halo WDs has been hampered by the lack of proper motion surveys that go deep enough
to find the cool halo WDs. The initial claims for a significant population of halo WDs in the field \citep{oppenheimer01a}
and in the Hubble Deep Field \citep{ibata00,mendez00} were later rejected by detailed model atmosphere analysis
\citep[see][and references therein]{bergeron05} and additional proper motion measurements \citep{kilic04,kilic05}.
To date, the coolest known probable halo WDs are WD 0346+246 \citep{hambly97,bergeron01} and SDSS J1102+4113,
with $T_{\rm eff} \approx 3800$ K \citep{hall08}. There are also about a dozen ultracool WDs detected in the SDSS
\citep{gates04,harris08} that may be thick disk or halo WDs, but current WD atmosphere models have problems in reproducing their
intriguing spectral energy distributions (SEDs). Therefore, their temperatures and ages remain uncertain.

Here we report the identification of three old halo WD candidates discovered as part of our Bok and USNO proper motion survey.
The details of this survey and our follow-up observations are discussed in Section 2, whereas our model fits and
analysis are discussed in Section 3. 

\section{Observations}

In January 2006, we started an $r-$band second-epoch astrometry survey
with the Steward Observatory Bok 90-inch telescope with its 90Prime camera \citep{williams01,liebert07}.
The 90Prime provides a field of view of 1.0 square degree with 0.45$\arcsec$ pixel$^{-1}$ resolution.
Since 2009 additional observations have been obtained with the U.S.
Naval Observatory Flagstaff Station 1.3m telescope using the CCD Mosaic Camera
(1.4 square degree field of view with 0.6$\arcsec$ pixel$^{-1}$ resolution).
We limited our program to the SDSS Data Release 3 footprint in order to have a relatively long time-baseline
between our program and the SDSS observations. We obtain proper motion errors of roughly 20 mas yr$^{-1}$
at $r=21$ mag ($g=22$ mag for cool WDs).

We select candidates for follow-up spectroscopy based on our proper motion measurements and the photometric colors.
We further limit our sample to objects with high proper motion and relatively red colors in order to find the elusive thick
disk and halo WDs. We started the follow-up optical spectroscopy of candidate halo WDs at the 6.5m MMT equipped with the Blue
Channel Spectrograph in June 2009. Here we present low resolution spectroscopy of three halo WD candidates
with $g-i=1.5-1.75$ mag. These observations were performed on UT 2009 June 19-21. Our targets are
SDSS J213730.87+105041.6,
J214538.16+110626.6, and J214538.60+110619.0 (hereafter J2137+1050, J2145+1106N, and J2145+1106S, respectively).
We used a 1.25$\arcsec$ slit and the 500 line mm$^{-1}$ grating
in first order to obtain spectra with wavelength coverage $3660-6800$ \AA\ and a resolving power of $R=$ 1200.
The $g-$band magnitudes of our targets range from 21.0 to 21.8 mag, and the exposure times range from 60 to 100 min.
We obtained all spectra at the parallactic angle and acquired He--Ar--Ne comparison lamp exposures for wavelength
calibration. We use the observations of the spectrophotometric standard star G24-9, which is also a cool WD, for
flux calibration.

In addition,  we obtained $J-$ and $H-$band imaging observations of our targets using the
MMT and Magellan Infrared Spectrograph \citep[MMIRS;][]{mcleod04} on the MMT on UT 2009
Sep 2 and 4. The FWHM of the images range from 0.8$\arcsec$ to 1.3$\arcsec$. We use a 1.0$\arcsec$ or 1.4$\arcsec$
aperture for photometry. The $6.8\arcmin \times 6.8\arcmin$ field of view of MMIRS enables us to use
20-50 nearby 2MASS stars to calibrate the photometry. The optical and near-infrared photometry of our
targets, as well as proper motions, are presented in Table 1. The optical photometry is in the AB system
and the $JH$ photometry is in the 2MASS (Vega) system. We use the corrections given in \citet{eisenstein06}
to convert the SDSS photometry to the AB system. Two of our targets, J2145+1106N and S (N-for
North and S-for South), are separated by 10$\arcsec$ and they have proper motions consistent within the errors.
Hence, they are in a common proper motion binary system.

\begin{deluxetable}{lccc}
\tabletypesize{\footnotesize}
\tablecolumns{4}
\tablewidth{0pt}
\tablecaption{New Halo White Dwarf Candidates}
\tablehead{
\colhead{Parameter}&
\colhead{J2137+1050}&
\colhead{J2145+1106N}&
\colhead{J2145+1106S}
}
\startdata
R.A.$^\dagger$                 & 21:37:30.87       & 21:45:38.16       & 21:45:38.60      \\
Dec.$^\dagger$                 & +10:50:41.6       & +11:06:26.6       & +11:06:19.0      \\
$\mu_{\alpha}$ (mas yr$^{-1}$) & $-$228.9          & +191.9            & +185.9           \\
$\mu_{\delta}$ (mas yr$^{-1}$) & $-$473.6          & $-$366.9          & $-$367.7         \\
$u$                            & 23.31 $\pm$ 0.69  & 23.74 $\pm$ 0.91  & 23.45 $\pm$ 0.75 \\
$g$                            & 21.77 $\pm$ 0.06  & 21.45 $\pm$ 0.05  & 21.00 $\pm$ 0.03 \\
$r$                            & 20.51 $\pm$ 0.03  & 20.27 $\pm$ 0.03  & 19.93 $\pm$ 0.02 \\
$i$                            & 20.02 $\pm$ 0.03  & 19.75 $\pm$ 0.02  & 19.49 $\pm$ 0.02 \\
$z$                            & 19.73 $\pm$ 0.08  & 19.68 $\pm$ 0.07  & 19.38 $\pm$ 0.06 \\
$J$                            & 19.21 $\pm$ 0.10  & 18.87 $\pm$ 0.07  & 18.54 $\pm$ 0.06 \\
$H$                            & 19.25 $\pm$ 0.18  & 19.00 $\pm$ 0.10  & 18.31 $\pm$ 0.06 \\
$T_{\rm eff}$ (K)              & 3780              & 3730              & 4110             \\
Age$^*$ (Gyr)                  & 9.6               & 9.7               & 8.7              \\
Distance$^*$ (pc)              & 78                & 69                & 70               \\
$V_{\rm tan}$ (km s$^{-1}$)    & 195               & 136               & 136              \\
$U, V, W$ (km s$^{-1}$)        & 172, $-$97, $-$35 & 31, $-$75, $-$102 & 31,  $-$75, $-$102
\enddata
\tablenotetext{\dagger}{Coordinates are given for equinox J2000.0 at the observed epoch of 2001.7.}
\tablenotetext{*}{These estimates are for $M=0.58~M_\odot$ ($\log g=8.0$). Ages, distances,
and velocities depend strongly on the assumed mass (see section 4.2).}
\end{deluxetable}

Out of the three targets, only J2145+1106S is detected in the USNO-B catalog, and it has a proper motion of $\mu_{\rm RA}=$
181.3 $\pm$ 5.2 mas yr$^{-1}$ and $\mu_{\rm DEC}= -367.7 \pm 5.2$ mas yr$^{-1}$
\citep{munn04}. These proper motion measurements are consistent with our measurements within the errors, and they
demonstrate that our proper motion measurements are reliable.

\section{Model Atmosphere Analysis}

Our MMT spectroscopy shows that all three targets have featureless spectra; they are cool DC WDs. 
Below 5000 K, H$\alpha$ disappears in cool WD spectra. However, hydrogen can still show its presence
through the red wing of Ly $\alpha$ absorption \citep{kowalski06b} in the blue and through collision-induced
absorption due to molecular hydrogen in the infrared \citep{hansen98,saumon99}. Cool helium atmosphere WDs
do not suffer from these opacities, and they are expected to show SEDs similar to blackbodies \citep{kowalski07b}.
Therefore, ultraviolet and near-infrared data are crucial for determining the atmospheric composition of
cool WDs.

We use state of the art white dwarf model atmospheres to fit the optical and near-infrared photometry
of our targets. The model atmospheres include the Ly $\alpha$ far red wing opacity \citep{kowalski06b}
as well as non-ideal physics of
dense helium that includes refraction \citep{kowalski04}, ionization equilibrium \citep{kowalski07b}, and the
non-ideal dissociation equilibrium of H$_2$ \citep{kowalski06a}.
Since parallax measurements are unavailable, we assume a surface gravity of $\log$ g = 8 ($M \approx 0.58 M_\odot$).
We discuss the implications of this mass assumption in section 4.2.

We find that the observed SEDs of our targets are best matched by pure hydrogen atmosphere models.
Figure 1 presents the observed and best-fit SEDs for our targets assuming a pure hydrogen atmosphere composition.
The temperatures for these models range from 3730 to 4110 K.
The SEDs peak around 1 $\mu$m. Even though the optical portion of the SEDs may be explained by simple blackbodies,
our $J-$ and $H-$band data show that they differ from blackbodies in the infrared.
The pure hydrogen atmosphere models match the ultraviolet, optical, and near-infrared SEDs of our targets fairly well. 

\begin{figure}
\plotone{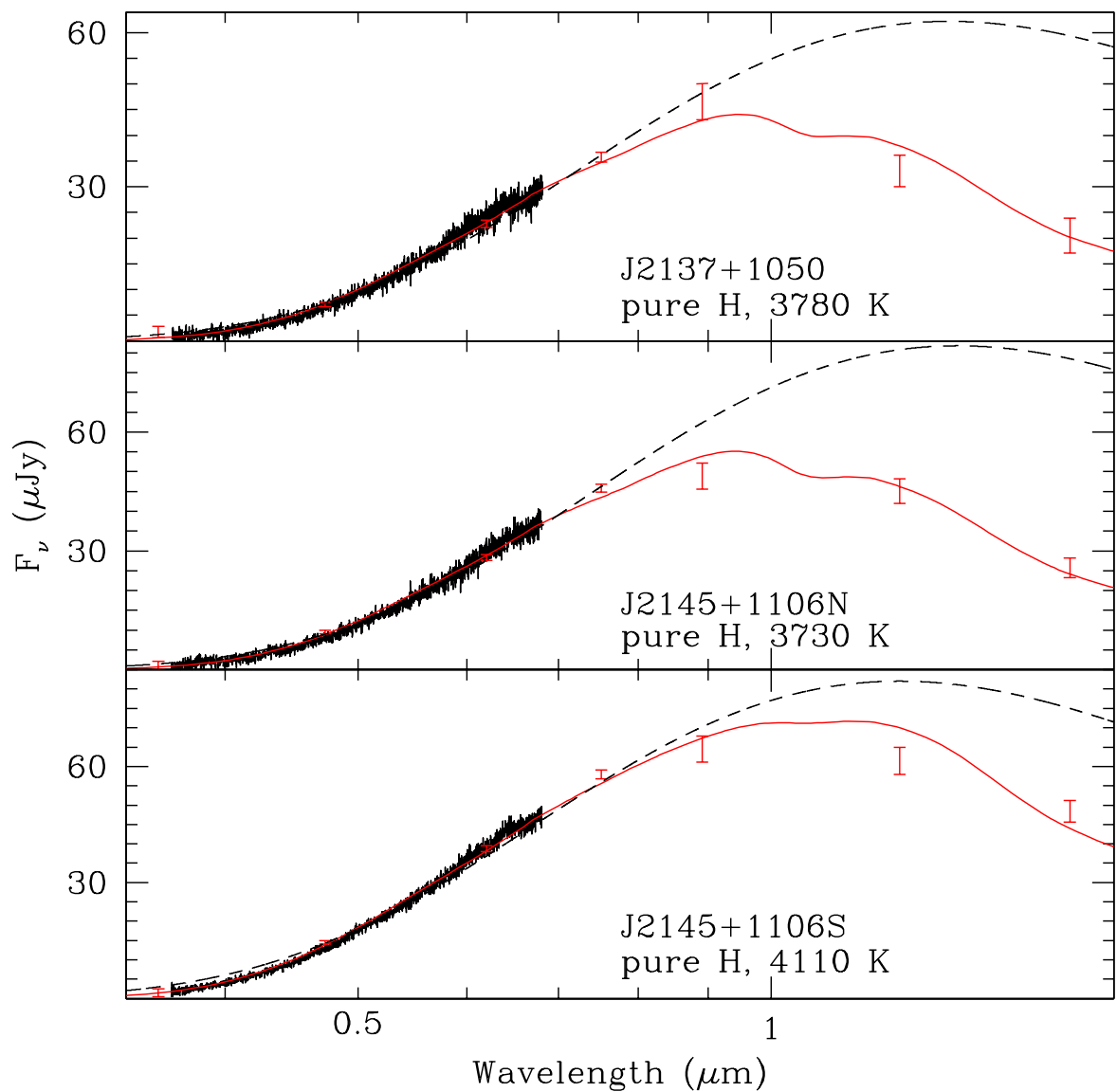}
\caption{Observed spectra, SDSS photometry, and near-infrared photometry of our targets compared to the
best-fit pure hydrogen atmosphere model spectra (solid lines, assuming $\log g=8$) and blackbody SEDs with
the same temperatures (dashed lines).}
\end{figure}

The fits for J2137+1050 and J2145+1106N are similar to the fits obtained for the halo WD candidate SDSS J1102+4113
\citep{hall08}. There are slight differences between the observations and these models.
The synthetic $i-$band fluxes seem lower than observed and there are related problems with matching
the $z-$ and $J-$ band fluxes. Systematic problems most likely exist for models below 4000 K.
The models for the two coolest stars predict absorption bumps around 1 $\mu$m. These bumps have never
been observed in the spectra of real WDs, indicating that the current collision-induced opacity calculations
may be problematic for high-density atmospheres of cool WDs \citep[see e.g.][]{oppenheimer01b,bergeron02,kilic09}.
Nevertheless, the overall SEDs of our targets agree with model predictions over the entire $0.3-1.7 \mu$m range.

Addition of helium can improve the model fits slightly. Figure 2 shows the best-fit models assuming pure H, pure He, 
and mixed H/He composition. Since mixed H/He atmospheres have higher pressure than pure H atmospheres at the same temperature,
the collision-induced absorption is expected to be stronger. Addition of 4-16\% helium (relative to hydrogen) into the atmosphere
helps with fitting the infrared portion of the SEDs. However, these fits are marginally better than the pure hydrogen atmosphere model
fits, and they are not statistically significant. The best-fit temperature values are also similar to the pure hydrogen
atmosphere solutions. Hence, the choice of a pure hydrogen or mixed H/He composition with small amounts of helium does not
significantly change our results. In any case, the good match between the optical spectrum and the models including
Ly $\alpha$ absorption indicates that these WDs have hydrogen-dominated atmospheres; helium-dominated or highly
helium-enriched atmospheres are ruled out \citep[see also][]{hall08}.

\begin{figure}
\plotone{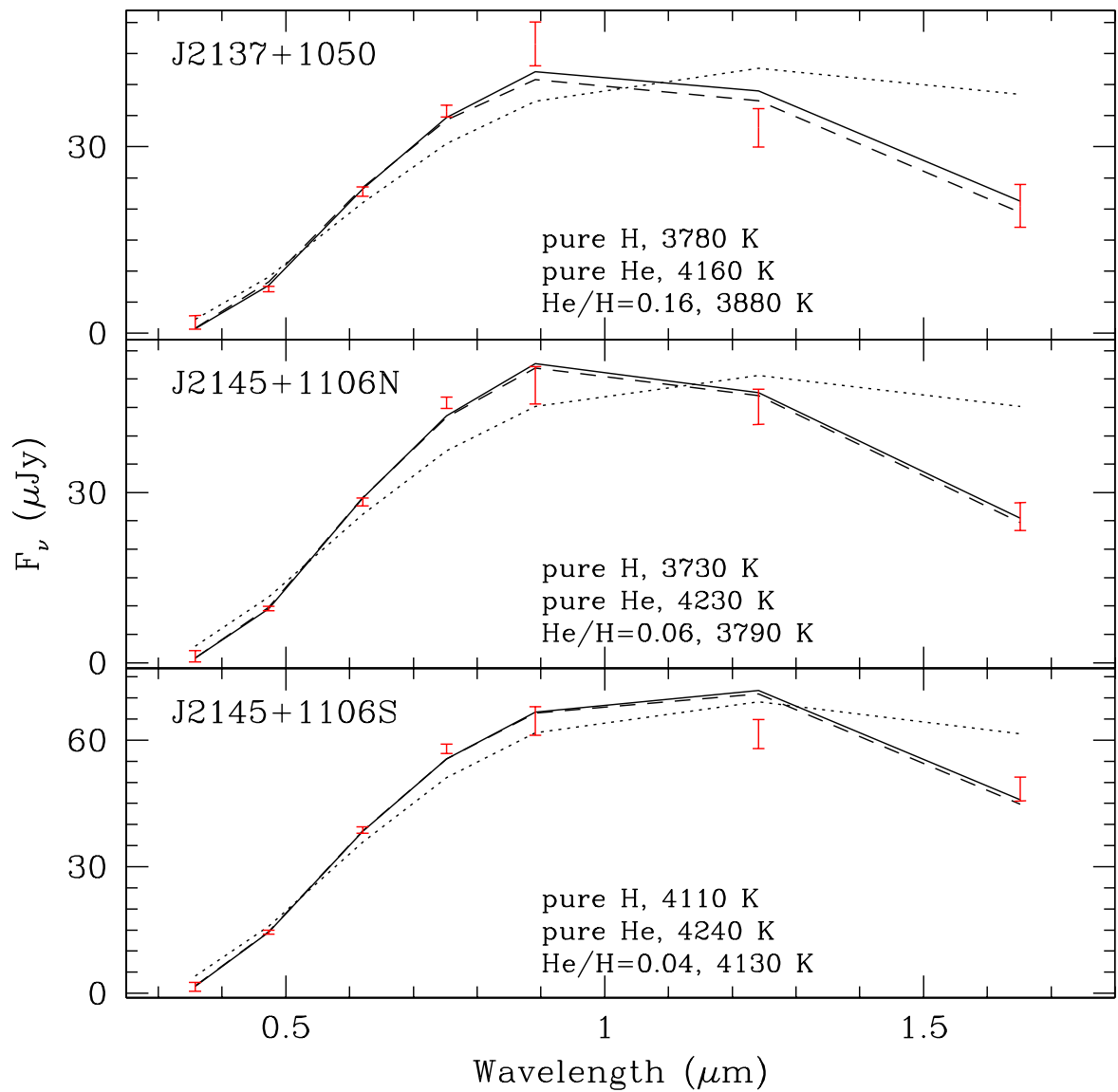}
\caption{A comparison of the observed photometry (error bars) with the synthetic photometry from the best-fit 
pure hydrogen (solid lines), pure helium (dotted lines), and mixed H/He (dashed lines) atmosphere models.
The best-fit temperatures are given in each panel.}
\end{figure}

The temperature, WD cooling age, and distance estimates for our targets based on pure hydrogen atmosphere
models with $\log g=8$ ($M=0.58 M_\odot$) and the cooling models by \citet{fontaine01} are given in Table 1.
This mass assumption implies that our targets
are located at 70-80 pc away from the Sun, and the WD cooling ages are 8.7 Gyr or longer.
Our model fits to the individual SEDs give cooling ages of 8.7-9.7 Gyr and distances of 69 and 70 pc for the members of the J2145+1106
common proper motion system. The difference in cooling ages can be explained by
a small mass difference between the two stars.
These results suggest that J2145+1106 system is a physical binary and that our model fits
are reliable. Based on our proper motion measurements and assuming zero radial velocity, we also estimate the
tangential velocity and Galactic $UVW$ velocities for our targets. These WDs display tangential velocities of 140-200 km s$^{-1}$.

\section{Thick Disk or Halo?}

\subsection{Total Ages}

\citet{bergeron05} emphasize the importance of determining total stellar ages in order to associate any WD with
thick disk or halo. Modelling the optical and near-IR SEDs of the \citet{oppenheimer01a} WD sample, \citet{bergeron05}
find that many of the WDs in that sample are fairly warm and too young to be halo WDs unless they all have masses near
0.5 $M_\odot$. They find that, with estimated temperatures of 3950-4100 K and ages of 8.8-9.1 Gyr, F351$-$50 and WD
0351$-$564 are the two most likely halo candidates in the \citet{oppenheimer01a} sample.

For an average mass of 0.58 $M_\odot$, our temperature estimates corresponds to WD cooling ages of 9.6-9.7 Gyr for
J2137+1050 and J2145+1106N. These two stars are the coolest field WDs currently known. Although the ultracool WDs discovered
by \citet{gates04} and \citet{harris08} are possibly cooler than our targets, current models have problems explaining 
the observed SEDs of these WDs \citep{bergeron02}.
Using the initial-final mass relations of \citet{williams09}, \citet{kalirai08}, and \citet{salaris09}, we estimate that
a 0.58 $M_\odot$ WD would be the descendant
of a $1.7-1.9 M_\odot$ star. Such a progenitor halo star has a main-sequence lifetime of
1.0-1.3 Gyr \citep{marigo08}. Therefore, the total ages of our 3 targets range from 9.7 to 11.0 Gyr;
they most likely belong to the halo or thick disk.
The theoretical uncertainties due to the unknown core composition, helium layer mass, crystallization, and phase separation
are on the order of 1 to 2 Gyr for these ages \citep[][M. Montgomery 2010, priv. comm.]{wood92,montgomery99}.

\subsection{Kinematic Membership}

Figure 3 shows the $UVW$ velocities of our targets, assuming they have 0 km s$^{-1}$ radial velocities,
compared to the 1$\sigma$ velocity ellipse of the halo and 2$\sigma$ ellipse of
the thick disk \citep{chiba00}. The velocities for the probable halo object WD 0346+246 are also shown for comparison.
The $U$ velocity of J2137+1050 is more than 3$\sigma$ different than the thick disk objects studied by
\citet{chiba00}. Similarly, the $W$ velocity of J2145+1106 is inconsistent with thick disk objects. The $UVW$ velocities
of the J2145+1106 binary are similar to that of WD 0346+246.
The radial velocity assumption does not change these results. Negative radial velocities bring the
$UVW$ velocities closer to the 1$\sigma$ distribution for the halo, and positive radial velocities move them
away from the 2$\sigma$ thick disk distribution (see the dashed lines in Fig. 3).
Hence, both J2137+1050 and J2145+1106 systems most likely belong to the halo.

\begin{figure}
\includegraphics[width=3.4in,angle=0]{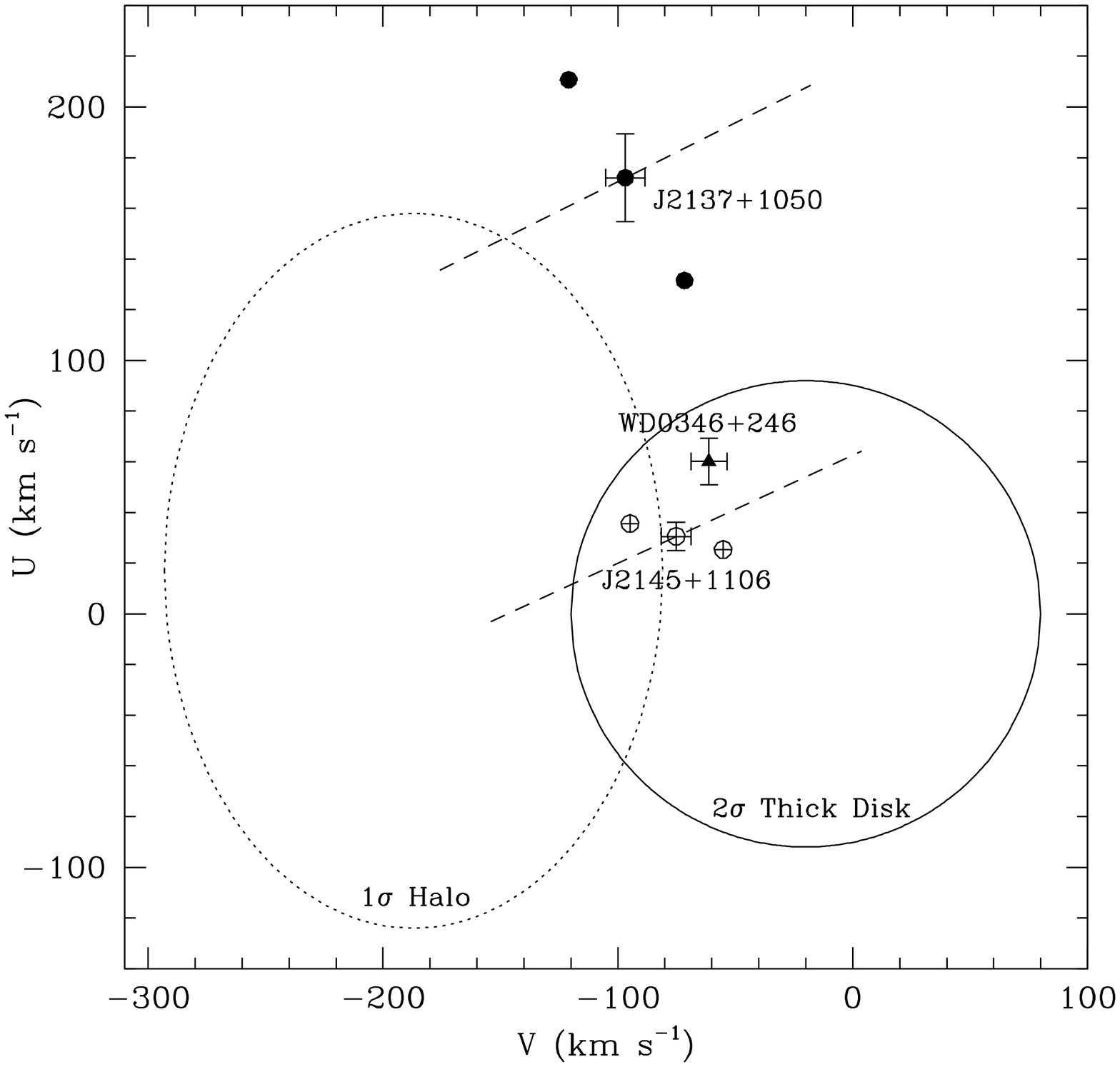}
\includegraphics[width=3.4in,angle=0]{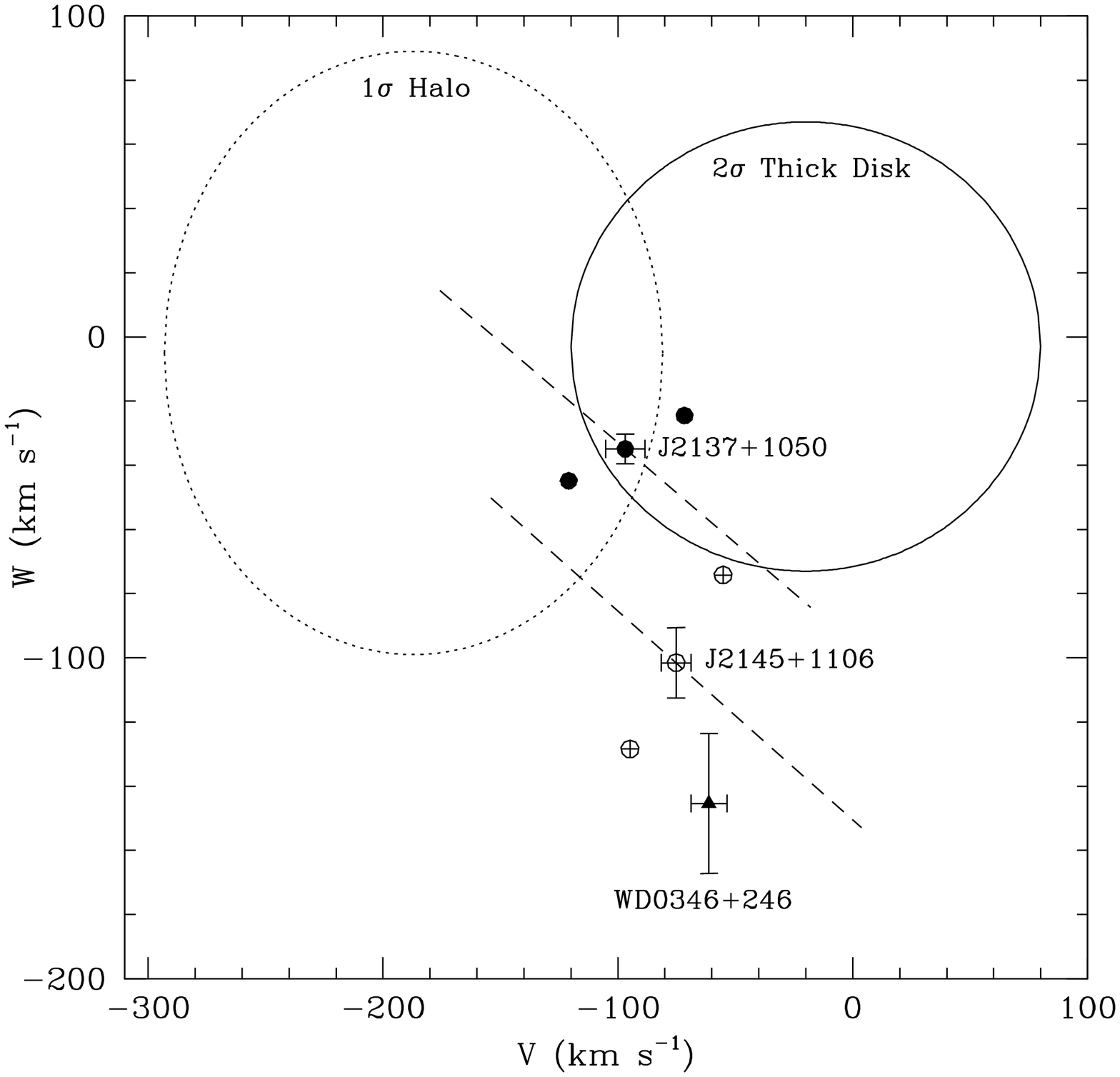}
\caption{$U,V$, and $W$ space velocities for our targets assuming 0 km s$^{-1}$ radial velocity and $\log~g=$ 7.5, 8.0, and 8.5 (from left
to right). The points with error bars correspond to $\log~g=$ 8. The probable halo member WD 0346+246 is shown
for comparison. The 2$\sigma$ velocity ellipse of the thick disk and the 1$\sigma$ ellipse of the halo are also shown.
The dashed lines show the effect of changing the radial velocity from $-$100 to +100 km s$^{-1}$ (from left to right).}
\end{figure}

Without a parallax measurement, our age, distance, and velocity estimates are of course uncertain. 
A $\log g$ of 8.5 ($M=0.9 M_\odot$) would imply WD cooling ages of 10.1-10.6 Gyr and $UVW$ velocities
that are still inconsistent with the 2$\sigma$ thick disk velocity distribution.
Likewise, a $\log g$ of 7.5 ($M=0.3M_\odot$) would imply WD cooling ages of 4.1-5.0 Gyr and $UVW$ velocities
that are even more inconsistent with the thick disk sample (see Figure 3). The main-sequence lifetimes would be greater
than the age of the universe unless the systems are unresolved double degenerates. 
An additional constraint is that J2145+1106
is a binary with a separation of 10$\arcsec$ (700 AU, assuming $\log g=8$).
This separation is too large to cause any effect on the evolution of each component and it is
small enough that the system can survive the gravitational perturbations from passing stars or Galactic tides
for billions of years (\citet{jiang10} demonstrate that more than 99.9\% of the binary stars with initial
separations of 0.017 pc ($\approx3500$ AU) survive for a Hubble time).
A scenario involving low-mass
WDs would require both components of the J2145+1106 system to be double degenerates, which seems unlikely.
In addition, our pure hydrogen atmosphere models with $\log g=8$ fit the SEDs better than the models with $\log g=7.5$,
indicating that our targets are not likely to be low-mass WDs.

Figure 4 displays a color magnitude diagram of the point sources in the region that encloses the WD population
of the globular cluster NGC 6397 and our three halo WD candidates assuming $\log g=$ 7.5, 8.0, and 8.5.
We use our best-fit WD model spectra to derive synthetic photometry in the $F606W$ and $F814W$ filters.
Depending on the mass, our targets can fall on multiple parts of the WD cooling sequence of NGC 6397.
If they are {\it similar} to the WDs in NGC 6397, they should have masses ranging from
0.5 $M_\odot$ to 0.9 $M_\odot$ ($\log g=7.9-8.5$).

\begin{figure}
\plotone{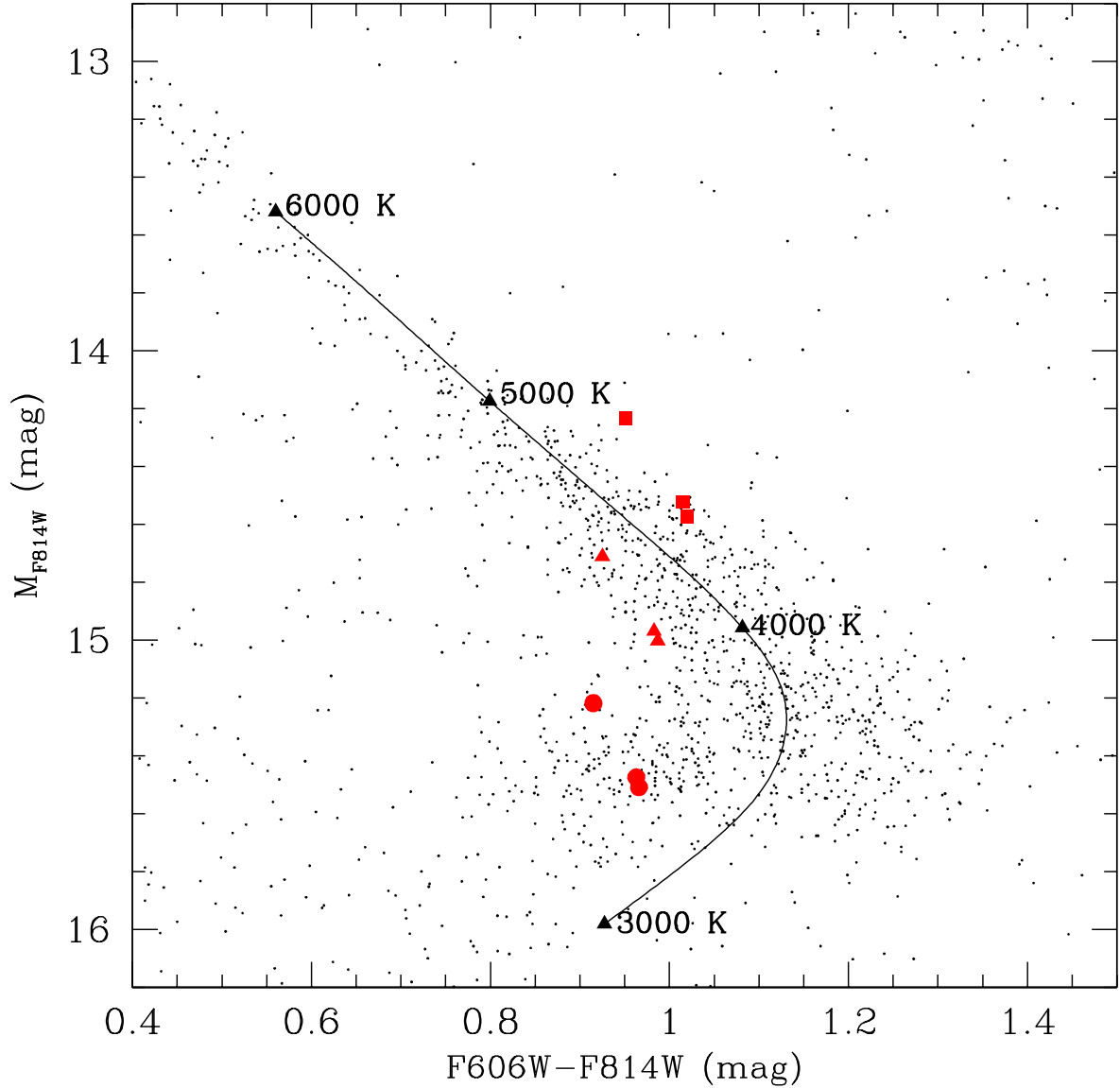}
\caption{A color-magnitude diagram of the WDs in the globular cluster NGC 6397 \citep[small dots,][]{hansen07} and
our three halo WD candidates assuming $\log g=$ 7.5, 8.0, and 8.5 (from top to bottom).
The NGC 6397 WD sequence is de-reddened by $E(F606W-F814W)=0.16$ and vertically shifted by
$\mu=12.0$ mag \citep[][]{kowalski07a}. The errors in reddening and distance modulus are on
the order of 0.03 and 0.06 mag, respectively.
The solid line shows the colors for 0.53 $M_\odot$
WDs with $T_{\rm eff}=3000-6000$ K.}
\end{figure}

\citet{kalirai09} find the masses of the brightest WDs in the globular cluster M4 to be 0.53 $M_\odot$.
This is a reasonable lower limit for our targets assuming single star evolution.
At 0.53 $M_\odot$, our targets would have WD cooling ages of 8.0-9.1 Gyr, distances
of 72-81 pc, and progenitor masses of 1.25-1.48 $M_\odot$ \citep{williams09,kalirai08}.
The main sequence lifetimes would be 1.7-2.8 Gyr for the progenitor halo stars \citep{marigo08}, and
the total ages would be 9.7-11.9 Gyr. The Galactic space velocities would be inconsistent with the thick disk velocity distribution.

\section{Conclusions}

J2137+1050 and J2145+1106 are cool WDs with hydrogen-dominated atmospheres. Our effective temperature estimates of 3730-3780 K
make J2137+1050 and J2145+1106N the coolest WDs known in the Solar neighborhood. Our best-fit models imply total ages of
$\approx 10-11$ Gyr, distances of 70-80 pc, and Galactic space velocities that are inconsistent with thick disk population within
$2\sigma$. We conclude that these targets most likely belong to the halo. However, trigonometric parallax observations
are required in order to constrain the distances, masses, and ages of our targets accurately. Such observations are currently
underway at the MDM 2.4m telescope.

Like WD 0346+246 and SDSS J1102+4113 \citep{bergeron01,hall08}, our three halo WD candidates have hydrogen-rich atmospheres.
The oldest WDs are likely to accrete from the interstellar medium within their $\sim$10 Gyr lifetimes and end up
as hydrogen-rich WDs even if they start with a pure helium atmosphere. However, the current
sample of halo WD candidates is not large enough to conclude that
most or all of the oldest WDs are hydrogen-rich. Observations of larger samples
of field WDs will be necessary to check whether all WDs turn into hydrogen-rich atmosphere WDs or not \citep[see the discussion
in][]{kowalski06b}.

The three targets that we present here make up only a small fraction of the halo WD candidates in our
proper motion survey. Follow-up observations of these targets will be necessary to confirm many more halo WD candidates
that can be used to study the age and age dispersion of the Galactic thick disk and halo. Already we can see, however,
that these halo (or possibly thick disk) WDs indicate a gap of 1--2 Gyr between the star formation in the halo and the
star formation in the disk at the solar annulus. Our observations further demonstrate that
deep, wide-field proper motion surveys ought to find many old halo WDs. 
Using the \citet{liebert88} WD luminosity function for the Galactic thin disk and a
single burst 12 Gyr old population with 10\% and 0.4\% local normalization for
the thick disk and halo, we estimate that there are 3200 thick disk and 140 halo WDs
per 1000 square degree (for a Galactic latitude of 45$^\circ$) down to a limiting magnitude of $V=21.5$ mag (our survey limit).
Pushing the limiting magnitude down to $V=24$ mag and assuming 50\% sky coverage, we estimate
that future surveys like the Pan-STARRS and LSST will image $\sim$ 1.3 million thick disk and $\sim$ 80,000 halo
WDs. These surveys will be invaluable resources for halo WD studies.

\acknowledgements
Support for this work was provided by NASA through the Spitzer Space Telescope Fellowship Program,
under an award from Caltech. This material is also based on work supported by the National Science
Foundation under grants AST-0607480 and AST-0602288. We thank the MMIRS commissioning team for obtaining the near-infrared
observations, and J. Liebert for extensive help with the Bok telescope imaging observations
and for many useful discussions. We also thank E. Olszewski for building the 90Prime
instrument and the Steward Observatory Time Allocation Committee for supporting our
proper motion survey program.

{\it Facilities:} \facility{MMT (Blue Channel Spectrograph and MMIRS), Bok (90 Prime)}

\end{document}